\documentclass[preprint,aps]{revtex4}

\usepackage{graphicx,amstext}

\begin{document}

\def \asf {$\mathrm{AsF_6}$\,}
\def \br {$\mathrm{Br}$\,}
\def \clo {$\mathrm{(TMTSF)_2ClO_4}$\,}
\def \pf {$\mathrm{(TMTSF)_2PF_6}$\,}
\def \spf {$\mathrm{(TMTTF)_2PF_6}$\,}
\def \tmtsfx {$\mathrm{(TMTSF)_2X}$\,}
\def \tmttfx {$\mathrm{(TMTTF)_2X}$\,}
\def \tmttfbr {$\mathrm{(TMTTF)_2Br}$\,}
\def \tmx {$\mathrm{(TM)_2X}$\,}
\def \tmtsf {$\mathrm{TMTSF}$\,}
\def \tmttf {$\mathrm{TMTTF}$\,}
\def \dcnqi {$\mathrm{DCNQI}$\,}
\def \bedt {$\mathrm{BEDT-TTF}$\,}
\def \didcnqix {$\mathrm{(DI\textsc{-}DCNQI)_2X}$\,}
\def \didcnqiag {$\mathrm{(DI\textsc{-}DCNQI)_2Ag}$\,}
\def \taf {$\mathrm{T_{AF}}$\,}
\def \tsc {$\mathrm{T_{SC}}$\,}
\def \edtasf {$\mathrm{[EDT\textsc{-}TTF\textsc{-}CONMe_2]_2AsF_6}$\,}
\def \edta {$\mathrm{EDT_{2}AsF_6}$\,}
\def \edtb {$\mathrm{EDT_{2}Br}$\,}
\def \edt {$\mathrm{EDT}$\,}
\def \edtbr {$\mathrm{[EDT\textsc{-}TTF\textsc{-}CONMe_2]_2Br}$\,}
\def \edtmol {$\mathrm{EDT\textsc{-}TTF\textsc{-}CONMe_2}$\,}
\def \edtx {$\mathrm{[EDT\textsc{-}TTF\textsc{-}CONMe_2]_2X}$\,}
\def \edt2x {$\mathrm{EDT_{2}X}$\,}

\def \alphai {$\mathrm{\alpha-(BEDT\textsc{-}TTF\textsc{-})_2I_3}$\,}
\def \kappaBEDT {$\mathrm{\kappa-(BEDT\textsc{-}TTF\textsc{-})_2X}$\,}
\def \H {$\mathrm{^1H}$\,}
\def \HO {$\mathrm{H_0}$\,}
\def \T1 {$\mathrm{T_{1}}$\,}
\def \C13 {$\mathrm{^{13}C}$\,}
\def \pstar {$P^{\star}$\,}
\def \Trho {$T_{\rho}$\,}
\def \1/T1 {${1/T_{1}}$\,}

\title{ Phase diagram of quarter-filled band organic salts, \edtx , \\  X =\asf  and  \br}

\author{P. Auban-Senzier}
 \affiliation {Laboratoire de Physique des Solides, UMR 8502, CNRS - Universit\'e Paris-Sud, B\^at. 510,  91405,Orsay, France}
\author{C. R. Pasquier}
\affiliation {Laboratoire de Physique des Solides, UMR 8502, CNRS - Universit\'e Paris-Sud, B\^at. 510,  91405,Orsay, France}
\author{D. J\'erome}
\affiliation {Laboratoire de Physique des Solides, UMR 8502, CNRS - Universit\'e Paris-Sud, B\^at. 510,  91405,Orsay, France}
\author{S. Suh}
 \affiliation {Department of Physics and Astronomy, University of California, Los Angeles, California 90095, USA}
 \author{S. E. Brown}
 \affiliation {Department of Physics and Astronomy, University of California, Los Angeles, California 90095, USA}
 \author{C. M\'ezi\`ere}
 \affiliation {Laboratoire Chimie, Ing\'enierie Mol\'eculaire et Mat\'eriaux, UMR 6200, CNRS-Universit\'e d'Angers, B\^at. K, 49045 Angers, France}
\author{P. Batail}
 \affiliation {Laboratoire Chimie, Ing\'enierie Mol\'eculaire et Mat\'eriaux, UMR 6200, CNRS-Universit\'e d'Angers, B\^at. K, 49045 Angers, France}

\begin{abstract}
An investigation of the $P/T$ phase diagram of the quarter-filled organic conductors, \edtx is reported on the basis of  transport and NMR studies of two members, X=AsF$_6$ and Br of the family. The strongly insulating character of these materials in the low pressure regime has been attributed to a remarkably stable charge ordered state confirmed by \C13 NMR and the only existence of 1/4 Umklapp e-e scattering favoring a charge ordering instead of the 1D Mott localization  seen in \tmx which are quarter-filled compounds with dimerization. A non magnetic insulating phase instead of the spin density wave state is stabilized in the deconfined regime of the phase diagram. This sequence of phases observed under pressure may be considered as a generic behavior for 1/4-filled conductors with correlations.
\end{abstract}

\pacs{74.70.Kn Organic superconductors}

\date{\today}

\maketitle

Superconductivity is just one of many phases exhibited by the \tmx series of organic salts (where $\mathrm{TM}$\  stands for  \tmttf  or \tmtsf  donor molecules) and for which correlations play a central role. This is illustrated by their phase diagram studied under pressure or magnetic field \cite{Springerreview,Brown08}. The fact that most members of the \tmttfx series behave as insulators under ambient pressure is of interest since on the basis of chemistry, structure and stoichiometry, conducting behavior is expected from the existence of partially filled bands.

The insulating ground states originate with electronic correlations, which take on an enhanced importance in the case of the quasi-one dimensional nature of the transfer integrals that applies here: ($t_a:t_b:t_c$=200meV:12meV:1meV)\cite{Jerome94}. For a 2:1 stoichiometry and monoanions, there is one hole in the highest occupied molecular orbital of every other molecules, and the conduction band deriving from the overlap of the wave functions along the stacking direction becomes one quarter empty with holes (or equivalently, three quarter filled with electrons). Hence, it is the 1/4-filled Umklapp scattering which is relevant\cite{Mila93}. Once an intra-stack dimerization is taken into account, two routes to insulating ground states are possible, and both are observed in the TMTTF compounds. For example, including the dimerization in the g-ology approach to the physics of 1D conductors \cite{Emery82} results in a localization of charges on the  bonds between adjacent molecules due to 1/2-filled Umklapp scattering\cite{Giamarchi04}. The ground state is referred to as a dimer-Mott insulator \cite{Barisic81}; in the real materials, spin-Peierls (SP) or antiferromagnetic (AF) ordering occurs at low temperature\cite{Bourbonnais99}. However, experiments reveal also a 4$k_F$ modulated charge density \cite{Chow00,Monceau01}, which is stabilized by strong near-neighbor Coulomb repulsion \cite{Seo97,Clay03}. Such a charge ordered (CO) configuration tends to favor an AF ground state, but here the exchange integrals, and consequently the AF wavevector, are controlled by the CO order parameter.

The details of the $T/P$ phase diagram for the TMTTF compounds clearly involve a mutual competition between these two effects \cite{Yu04} and the kinetic energy \cite{Jaccard01}. The opportunity to study the progression insulator$\to$metal in a truly 1/4-filled quasi-one dimensional compound ({\it i.e.}, without dimerization) presents itself with the crystallization of \edtx, for which there is no center of symmetry between the molecules along the stack, at variance with the \tmx series and their inherent dimers, and where extensive refinement of X-ray synchrotron data indicate uniform distribution of the molecular units, yielding a band quarter-filled with holes \cite{Xray}.

\begin{figure}[ht]
\centerline{\includegraphics[width=0.8\hsize]{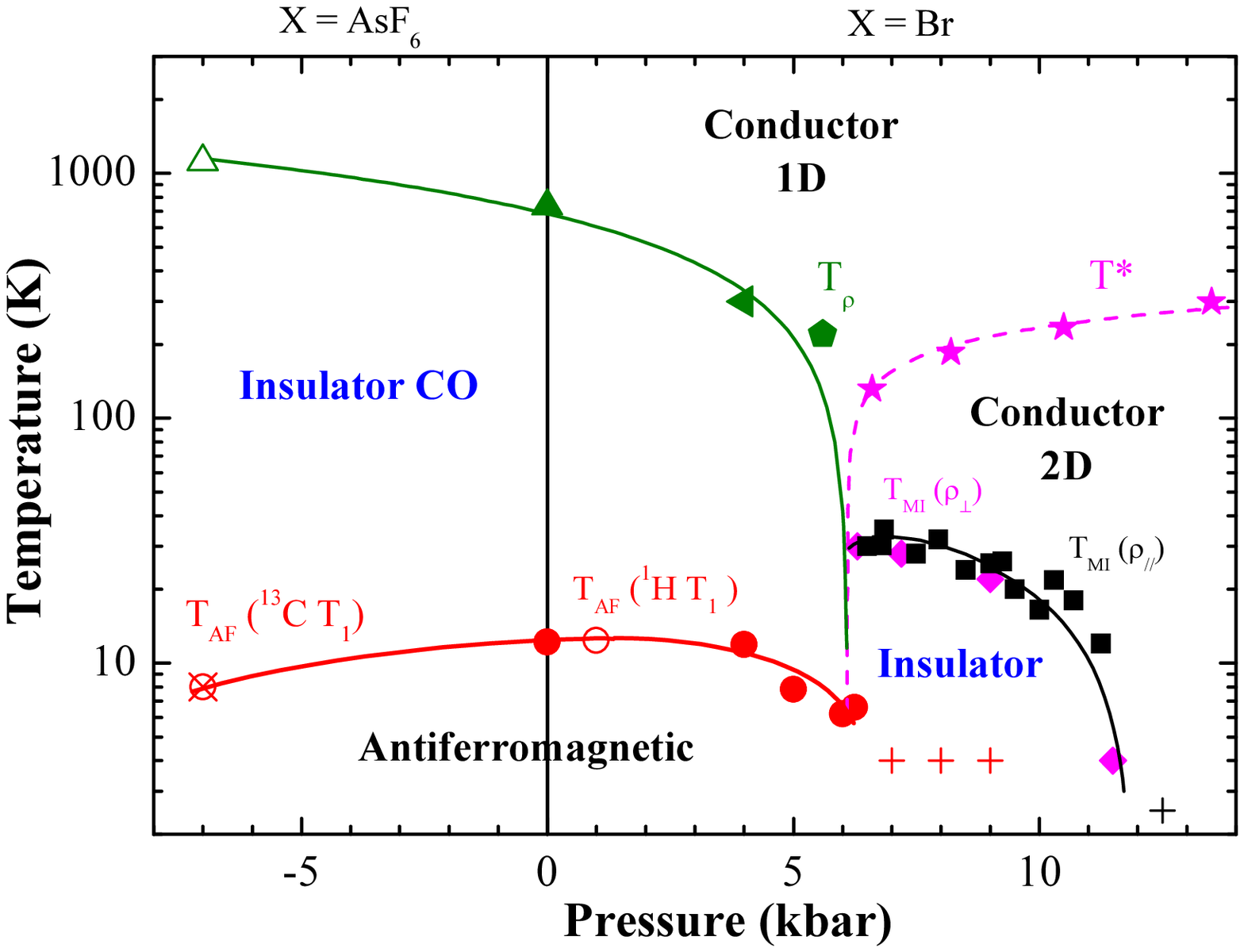}}
\caption{Generic ($P/T$) phase diagram of \edtx with open and close symbols for \edta and \edtb respectively. The origin of pressure is taken for \edtb. High temperature triangles provide a determination of $T_{CO}$ from the $P$-dependence of the longitudinal conductivity at room temperature and a BCS relation between the gap and the transition temperature. $T_{CO}$ is also deduced from the kink on the $P$-dependence of the conductivity at room temperature (triangle at 4 kbar). The minimum of the longitudinal resistivity ($T_{\rho}$) in the 1D regime is reported when it is observed below room temperature (at 5.6 kbar). Circles (cross (x) indicate the  antiferromagnetic ground state observed by \H (\C13 ) NMR.   Crosses at 4 K and 7 - 9 kbar mean the lowest temperature reached without any magnetic ordering. Squares indicate the metal to insulator transition derived from transverse and longitudinal resistivity.The cross at (3K, 12.5 kbar) marks the onset of a weak localisation also present at higher pressure.}
\label{DPlog_EDTAsBr2.pdf}
\end{figure}

\begin{figure}[hb]
\centerline{\includegraphics[width=0.8\hsize]{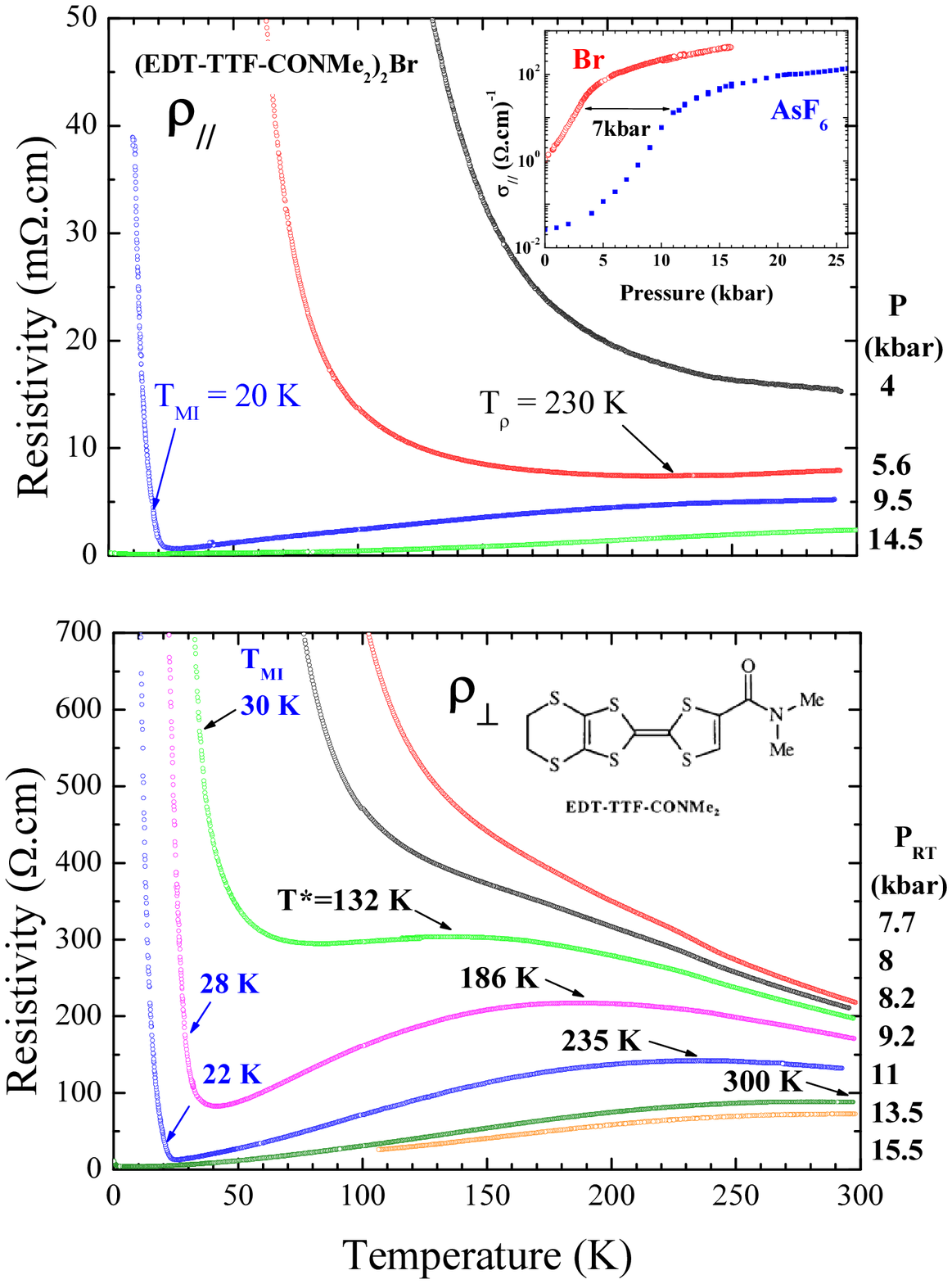}}
\caption{Temperature dependence of longitudinal (upper panel) and  transverse resistivity (lower panel) of \edtb for different  pressures. The upper insert shows the $P$-dependence of the longitudinal conductivity of \edtb and \edta with kinks at 4 and 11 kbar respectively displaying a cross over from an exponential to a linear $P$-dependence. }
\label{RTPparaperp.pdf}
\end{figure}

This report describes the effect of high pressure on the physical properties of two isostructural members of the family, namely X = \br and \asf (\edtb and \edta), where the smaller volume of the bromide ions plays the role of a chemical pressure, in the same way as is already known from the study of the \tmx series. The $T/P$ phase diagram shown in Fig.\ref{DPlog_EDTAsBr2.pdf}, and inferred from a combination of transport and NMR measurements, is simpler than for the TMTTF systems because of the absence of dimerized stacks. At ambient pressure and $T$=300K, both compounds are in a CO (Mott) insulating state and cooling results in antiferromagnetically ordered ground states. Application of pressure suppresses the CO phase in favor of a quasi-one dimensional conductor that, upon cooling, undergoes a finite temperature transition that is likely a Peierls instability. Still higher pressures suppress the Peierls transition, presumably as the nesting condition is weakened. Note that similar properties were observed for both compounds once the pressure is shifted for the AsF$_6$ material by -7 kbar relative to \edtb (see the inset of Fig.\ref{RTPparaperp.pdf}, top panel), and for this reason the phase diagram as presented applies to both compounds, \edtb and \edta, \textit{albeit} with the appropriate shift in pressure. Experiments on another structurally more complex 1/4-filled system, \didcnqix \cite{Itou04,Kakiuchi07} have not yet clarified the existence of an intermediate density wave insulator.

Single crystals of \edta and \edtb have been prepared following the procedure described in refs.\cite{Xray,Heuze03}. Resistivity measurements under pressure and at low temperature have been conducted on single crystal needles either in longitudinal or transverse configurations using standard techniques also described in ref.\cite{Heuze03}.

\H and $^{13}$C NMR experiments have been carried out on single crystals of \edta and \edtb, with the stacking axis \textit{a}$\perp$\HO. For the case of $^1$H nuclei, the field strength was \HO=2-2.4T, and the spin-lattice relaxation time $T_{1}$ has been measured in the temperature range 4.2-40 K, where only the electronic contribution to the relaxation is present. For the \C13 experiments, one carbon site (C1) was 100\% spin labelled, and \HO=10T. Hydrostatic pressure was applied using CuBe and NiCrAl clamp cells and Daphne 7373 silicone oil (transport) and Flourinert ($^1$H NMR) as the pressure transmitting medium. Ambient temperature pressures were determined using a previous calibration based on a manganin resistance gauge. For all experiments, 2 kbar are removed at low temperature to account for the loss of pressure during cooling.

In Fig.\ref{RTPparaperp.pdf} are shown longitudinal resistivity data $\rho_{\parallel}(T)$ (top panel) of \edtb for various pressures, and the same for the transverse direction, $\rho_{\perp}(T)$ (lower panel). From these results, and what is known previously for \edta \cite{Heuze03}, the resistivity components are activated for all temperatures less than 300K. Some qualitative changes are evident for $\rho_{\perp}$ at pressures beyond $P\sim$7 kbar. First, a wide maximum in the transverse resistivity emerges at intermediate temperatures; it is marked as $T^*$ in Fig. \ref{DPlog_EDTAsBr2.pdf}. Such a maximum can be related to the deconfinement of the 1D carriers lowering the temperature as already studied in-depth in the \tmx series\cite{Moser98}. Second, there is evidence for a phase transition to an insulating state at lower temperatures, marked as $T_{MI}$.

\begin{figure}[ht]
\centerline{\includegraphics[width=0.8\hsize]{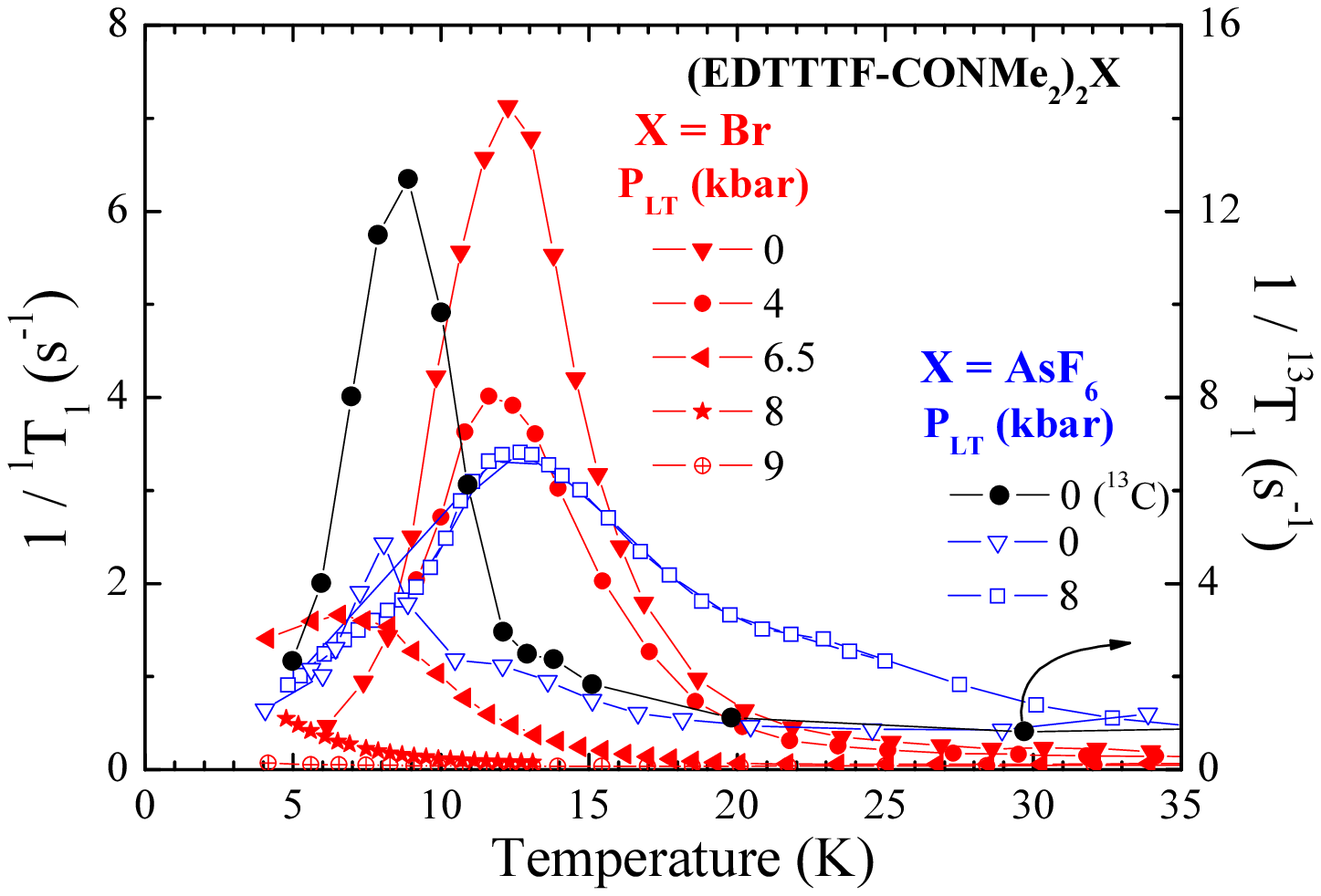}}
\caption{\H  and \C13 \1/T1  data of \edta and \edtb }
\label{invT1T_EDTAsF6Br.eps}
\end{figure}

The existence and pressure dependence of the transition to antiferromagnetic order is known from measurements of the $^1$H and $^{13}$C spin lattice relaxation rates, shown in Fig. \ref{invT1T_EDTAsF6Br.eps}. The peak appears similar to what is seen at the N\'eel temperature $T_N$ for other organic systems, such as (TMTTF)$_2$Br \cite{Creuzet82} or $\kappa$-(BEDT-TTF)$_2$Cu[N(CN)$_2$]Cl \cite{Lefebvre00}. Below this temperature, the NMR spectra broaden as expected for antiferromagnetic ordering (not shown). Note that for pressures $P>$7 kbar, no peak is observed in $^1$H $T_1^{-1}$ at any temperature, including $T_{MI}$.

\begin{figure}[hbt]
\includegraphics[width=0.89\hsize]{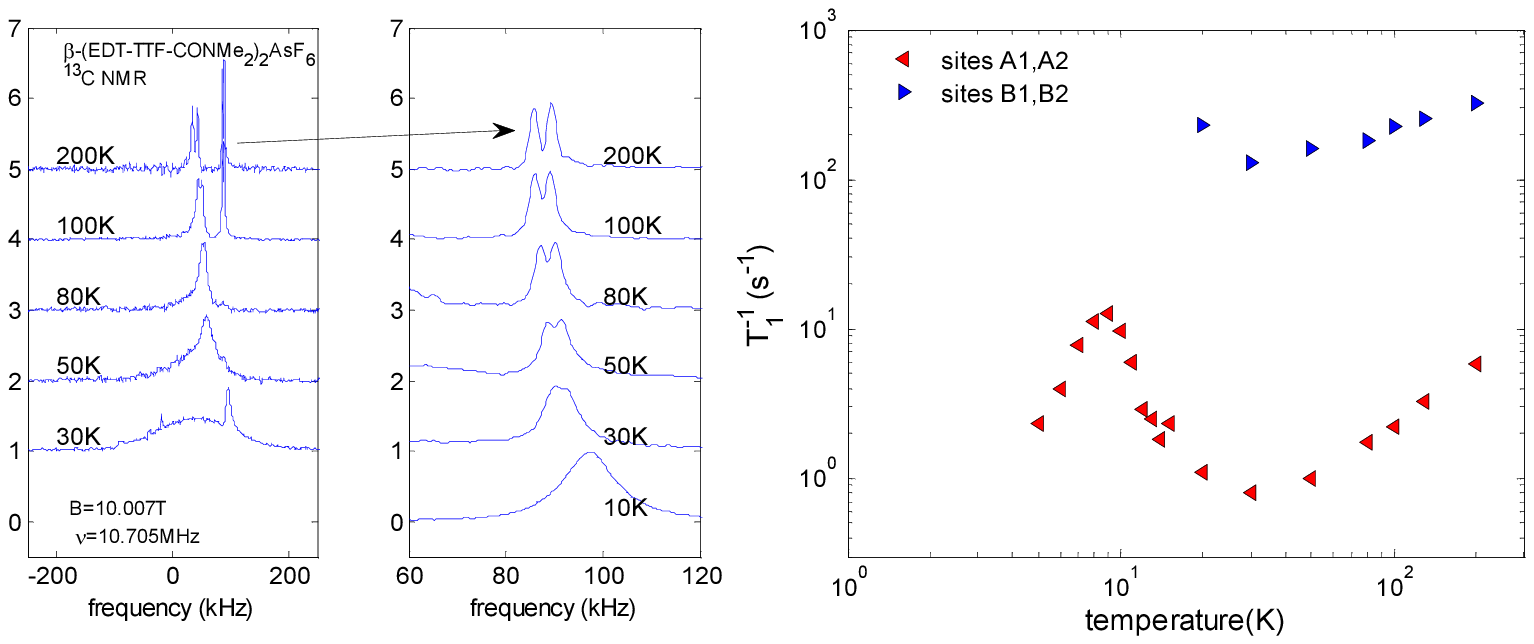}
\caption{Top left: Temperature evolution of the \C13  spectra for  \edta. The upper right-hand panel follows the pair of lines with the weaker hyperfine coupling called A sites for molecular orientations 1 and 2. Lower: $1/T_{1} vs (T)$ . The B charge rich site spins relax at a rate about two orders of magnitude faster than the A charge depleted site spins.}
\label{Stuart 3.pdf}
\end{figure}

Finally, we comment on the evidence for charge ordering (CO) at low pressure, and the line of transitions separating the CO and high-symmetry phases. The Arrhenius plot of the resistivity of \edta shown in Ref. \cite{Heuze03} demonstrates that the activated regime is well established already at room temperature with an energy of $1350K$ at 1 bar. We can associate the charge gap with the CO using \C13 spectroscopy and relaxation measurements, and recent confirmation by X-ray diffraction \cite{Xray}. The \C13 \ NMR  spectrum of \edta\ recorded  at $T=200K$ and below, displays four absorption lines, Fig.\ref{Stuart 3.pdf} (upper panel). This is twice the number of lines expected for the two molecular orientations of the C1  atom in the structure. Consequently, this NMR spectrum provides the evidence that the \asf salt is in a CO phase: the doubling of the number of resolved peaks indicates that for each of the two molecular orientations 1 and 2, there are two distinct magnetic environments A and B due to a charge imbalance. The spin-lattice relaxation rate, $T_{1}^{-1} (T)$ versus $T$ for sites A, B is shown in Fig.\ref{Stuart 3.pdf}, (lower panel). The $T$-dependence of the relaxation rate, governed by uniform spin fluctuations in the high temperature domain is consistent with the data of \spf \cite{Bourbonnais99}.  The rates are different by nearly two orders of magnitude, $T_1^{-1}(A)\approx O(100\, T_1^{-1}(B))$ over the entire range of measured temperatures. A weak $T$-dependence of the order parameter (CO amplitude) can account for the $T$-dependence difference between sites A, B \cite{Zamborsky02}. The observation of very different rates is qualitatively similar to that observed in other charge transfer salts with CO symmetry breaking, such as \spf \cite{Chow00,Yu04}, \didcnqiag \cite{Hiraki98}. Following the discussions of, for example, the \tmttf materials, we consider the charge imbalance ratio as,
$\frac{T_{1}^{-1}(B)}{T_{1}^{-1}(A)}=\left(\frac{\rho_B}{\rho_A}\right)^2$ and consequently, the charge ratios for sites A,B is roughly 9:1 and essentially $T$ independent from 10 to 200K.

Evidence that pressure suppresses the CO ordering is shown in the inset to the top panel of Fig. \ref{RTPparaperp.pdf}, in which the pressure dependence of the longitudinal conductivity $\sigma_{\parallel}(P,T=300\text{K})$ exhibits a sharp change in behavior: the room temperature conductivity of \edta depends exponentially on pressure up to  11 kbar, (\textit{see} insert in fig.\ref{RTPparaperp.pdf}) while it  becomes linearly pressure-dependent above. A similar kink in the pressure dependence of $\sigma$ is observed for \edtb, although at $P\sim$4 kbar, and its existence is taken to mark the transition to the CO state at ambient temperature. Assuming that the gap vanishes at the pressure of the kinks together with the pressure dependence of the conductivity up to the kink pressure, we can derive an estimate of the transition temperature using a BCS-like relation between the gap and the transition temperature. Thus, the charge gap of \edta at ambient pressure would be stable even above $1000K$.

At high temperatures, the \edt2x and  \tmx phase diagrams are quite similar.  Namely, the existence of a wide pressure regime exhibiting 1D localization and a deconfinement border of the 1D carriers arising above a pressure of about 6 kbar as detected by the existence of a wide maximum in the $T$-dependence of $\rho_{\bot} (T)$ (Fig.\ref{RTPparaperp.pdf}). Furthermore, above the deconfinement pressure a sharp metal-insulator transition is observed for both components of the resistivity, Fig.\ref{RTPparaperp.pdf},  near $T\sim$30 K  with a strong  suppression under pressure. High pressure runs performed at 14.5 and 19.5 kbar in a dilution refrigerator  have revealed a metallic $T^2$ law behavior with a fairly large residual value ( $\rho_{\parallel,o}$= $125\pm 25(\mu\Omega .cm)$. No sign of superconductivity could be observed down to 70 mK.

The phase diagram shown in Fig.\ref{DPlog_EDTAsBr2.pdf} reveals important features. First, the CO phase, which is stable at low pressure is suppressed by a deconfinement pressure of  $P^\star=6$ kbar for \edtb. Second, the AF phase is also  suppressed in the vicinity of the critical pressure. Third, the high pressure insulating phase, which is stable at low temperature in the 2D regime, cannot be considered as a simple extension above \pstar of the AF phase existing below as no magnetic ordering could be detected up to the transition temperature \textit{via} $^1$H $T_{1}^{-1} (T)$ measurements. Since the metal-insulator transition occurring above deconfinement is strongly depressed under pressure we suggest that this transition is triggered by the nesting properties of the Fermi surface  (Peierls-like transition) in a way similar to the SDW state of \pf  and fully suppressed when the frustration of this nesting becomes dominant at pressures higher than 13 kbar.

In conclusion, the experimental study of the $P/T$ phase diagram of the newly synthesized series of 1/4-filled organic salts \edtx has revealed significant differences from the well established phase diagram of \tmx salts which are quarter-filled systems with intra-stack dimerization. In particular, localization due to 1/2-filled Umklapp in 1D dimer-Mott compounds is absent in 1/4-filled systems, and consequently the spin-Peierls ground state seen in \spf\ is also missing. At low pressures, Coulomb interactions extending beyond on-site lead to an insulating phase due to charge ordering. In this case, the  disproportionation ratio is very large, $\sim$9:1. Above the deconfinement pressure an insulating state is stabilized but rapidly suppressed under pressure allowing a metallic state to become stable  without any trace of superconductivity above 70mK. The absence of superconductivity in the pressure stabilized metallic state  may be related to the absence of any close proximity with  the magnetic phase or to some small amount of structural defects in the twinned low temperature monoclinic structure\cite{Xray}. Thus, the 1/4-filled system with correlations is characterized by a very simple, probably generic, sequence of phases: CO insulator$\to$density wave insulator$\to$Fermi Liquid. The phase diagram may be similar to the one of \didcnqiag\cite{Hiraki98}, characterizing 1/4-filled 1D conductors, although it is not yet clarified whether an intermediate density wave insulator exists in that case \cite{Itou04,Itou05}.

\acknowledgments
This work was supported by the Agence Nationale de la Recherche  under Grant ANR CHIRASYM 2005-08 (NT05-2 42710) . We also acknowledge P.Foury-Leylekian and J.P.Pouget for fruitful discussions and P. Wzietek for his cooperation. Work at UCLA was supported in part by the National Science Foundation under grant numbers DMR-0520552 and DMR-0804625.

\end{document}